\documentclass[twocolumn]{aastex61}
\usepackage{xcolor}

\received{April 10, 2018}
\revised{June 14, 2018}
\accepted{June 22, 2018}

\shorttitle{Mass--temperature relation of galaxy clusters}
\shortauthors{Fujita et al.}

\begin{document}

\title{A New Interpretation of the Mass-Temperature Relation and Mass Calibration of Galaxy Clusters Based on the Fundamental Plane}

\correspondingauthor{Yutaka Fujita}
\email{fujita@astro-osaka.jp}

\author[0000-0003-0058-9719]{Yutaka Fujita}
\affil{Department of Earth and Space Science, Graduate School
 of Science, Osaka University, Toyonaka, Osaka 560-0043, Japan}

\author{Keiichi Umetsu}
\affiliation{Institute of Astronomy and Astrophysics, Academia
Sinica, P.O. Box 23-141, Taipei 10617, Taiwan}

\author{Stefano Ettori}  
\affiliation{INAF, Osservatorio di Astrofisica e Scienza dello Spazio, via Pietro Gobetti 93/3, I-40129 Bologna, Italy}
\affiliation{INFN, Sezione di Bologna, viale Berti Pichat 6/2, I-40127 Bologna, Italy}

\author{Elena Rasia}
\affiliation{INAF, Osservatorio Astronomico di Trieste, via Tiepolo 11,
 I-34131, Trieste, Italy}

\author{Nobuhiro Okabe}
\affiliation{Department of Physical Science, Hiroshima University, 1-3-1
 Kagamiyama, Higashi-Hiroshima, Hiroshima 739-8526, Japan}

\author{Massimo Meneghetti} 
\affiliation{INAF, Osservatorio di Astrofisica e Scienza dello Spazio, via Pietro Gobetti 93/3, I-40129 Bologna, Italy}
\affiliation{INFN, Sezione di Bologna, viale Berti Pichat 6/2, I-40127 Bologna, Italy}

\begin{abstract}

Observations and numerical simulations have shown that the relation
between the mass scaled with the critical density of the universe and
the X-ray temperature of galaxy clusters is approximately represented by
$M_\Delta \propto T_X^{3/2}$ (e.g. $\Delta=500$). This relation is often
interpreted as evidence that clusters are in virial
equilibrium. However, the recently discovered fundamental plane (FP) of
clusters indicates that the temperature of clusters primarily depends on
a combination of the characteristic mass $M_s$ and radius $r_s$ of the
Navarro--Frenk--White profile rather than $M_\Delta$. Moreover, the
angle of the FP revealed that clusters are not in virial equilibrium
because of continuous mass accretion from the surrounding matter. By
considering both the FP and the mass dependence of the cluster
concentration parameter, we show that this paradox can be solved and the
relation $M_\Delta \propto T_X^{3/2}$ actually reflects the central
structure of clusters. We also find that the intrinsic scatter in the
halo concentration--mass relation can largely account for the spread of
clusters on the FP. We also show that X-ray data alone form the FP and
the angle and the position are consistent with those of the FP
constructed from gravitational lensing data. We demonstrate that a
possible shift between the two FPs can be used to calibrate cluster
masses obtained via X-ray observations.

\end{abstract}


\section{Introduction}
\label{sec:intro}

Clusters of galaxies are the most massive objects in the
universe. Because of their large scale, they have been expected to
reflect the properties of the background universe and have been used to
study cosmological parameters such as the amount of matter and dark
energy, and to study the growth of large scale structures. In
particular, the mass function is one of the most powerful tools for that
purpose
\citep[e.g.][]{1993ApJ...407L..49B,eke96a,rei02a,2009ApJ...692.1060V}. The
cluster masses have been measured via X-ray or gravitational lensing
observations. However, different measurements may be affected by
different biases, which makes it complicated to be compared with the
mass function from numerical simulations. To correct those biases,
statistical approaches have been considered useful. In particular,
scaling relations for a large number of clusters have been used as an
efficient tool to estimate the biases. The relation between the mass and
the X-ray temperature of the intracluster medium (ICM) or $T_X$ is a
representative one. For this relation, previous studies have adopted
$M_\Delta$ or the mass enclosed within a sphere of radius $r_\Delta$,
within which the mean overdensity equals $\Delta$ times the critical
density of the universe $\rho_c$. The critical density depends on
redshift $z$ as in $\rho_c(z)\propto E(z)^2$, where $E(z)$ is the Hubble
constant at $z$ normalized by the current value $H_0$. The values of
$\Delta=200$ and 500 have often been used. Observations and numerical
simulations have shown that the relation is approximately represented by
$M_\Delta\propto T_X^{3/2}$ for clusters
\citep{bry98a,ett02b,sun09a,2016A&A...592A...4L,tru18a}.

This relation has often been explained as follows. Assuming that the
representative density of clusters is $\rho_\Delta\equiv \Delta\rho_c(z)
\propto \Delta E(z)^2$, the density does not depend on the mass at a
given redshift.  If a cluster is isolated and well-relaxed or
``virialized'' inside $r_\Delta$, it is represented by a sphere of the
radius $r_\Delta$, which is close to isothermal. In this case, the mass
is given by $M_\Delta=4\pi\rho_\Delta r_\Delta^3/3$ and the temperature
is given by $T_X\propto M_\Delta/r_\Delta\propto \rho_\Delta r_\Delta^2
\propto \Delta E(z)^2 r_\Delta^3$. From these relations, the
mass--temperature relation can be represented by
\begin{equation}
\label{eq:MTclassic}
 M_\Delta \propto T_X^{3/2}\Delta^{-1/2} E(z)^{-1}\:.
\end{equation}
\citep[e.g.][]{kai86a,bry98a,2011ASL.....4..204B,2015SSRv..188...93P},
which is generally consistent with observations and simulations.

However, this conventional interpretation may appear to be at odds with
a concept that came from recent numerical simulations. According to
$N$-body simulations, the density profile of dark matter halos of galaxy
clusters is not the isothermal profile ($\propto r^{-2}$) but can be
represented by the Navarro--Frenk--White (NFW, hereafter) density
profile:
\begin{equation}
\label{eq:NFW}
 \rho_{\rm DM}(r) = \frac{\delta_c\rho_c}{(r/r_s)(1+r/r_s)^2}\:,
\end{equation}
where $r$ is the clustercentric distance, and $r_s$ is the
characteristic or scale radius \citep{nav97a}. The normalization of the
profile is given by $\delta_c$. We define the mass inside $r_s$ as
$M_s$. The ratio
\begin{equation}
 \label{eq:c}
c_\Delta=r_\Delta/r_s
\end{equation}
is called the concentration parameter and $c_\Delta>1$ for $\Delta=200$
and 500 for clusters. The ``inside-out'' formation scenario of galaxy
clusters has been proposed based on the results of the $N$-body
simulations \citep{sal98a,fuj99d,bul01a,wec02a,zha03a}. In this
scenario, the inner region ($r\lesssim r_s$) forms rapidly, and only
successively the outer region ($r\gtrsim r_s$) slowly grows through
matter accretion. The inner structure at $r\lesssim r_s$ established
during the fast-growing phase is well conserved in the latter
slow-growing phase. The cluster formation time can be defined as the
shift-time from the fast-growing phase to the slow-growing
phase. Clusters that formed earlier tend to have higher characteristic
density $\rho_s\equiv 3 M_s/(4\pi r_s^3)$
\citep{nav97a,zha09a,lud13a,cor15c}. Contrary to $\rho_\Delta$, this
representative density $\rho_s$ is not constant among clusters at a
given $z$. Moreover, the inside-out scenario indicates that clusters are
not well-relaxed in the entire region within $r_\Delta$. These suggest
that the conventional virial interpretation of the relation
$M_\Delta\propto T_X^{3/2}$ might need to be reconsidered.

In this study, we propose a new interpretation of the relation
$M_\Delta\propto T_X^{3/2}$, which is consistent with the inside-out
scenario. This interpretation is based on the newly discovered
fundamental plane (FP) of clusters (\citealp{fuj18a}, hereafter Paper~I)
and the mass dependence of the concentration parameter $c_\Delta$. We
also investigate the FP constructed from X-ray data alone in order to
endorse studies of scaling relations based on X-ray data. The rest of
the paper is organized as follows. In section~\ref{sec:FP}, we summarize
the properties of the FP. In section~\ref{sec:MT}, we show that the
mass--temperature relation can be explained by considering both the
FP and the cluster concentration parameter, $c_\Delta$. In
section~\ref{sec:XFP}, we show that the FP formed by X-ray data alone is
consistent with the FP we found in Paper~I. In section~\ref{sec:calib},
we propose a new calibration method of cluster mass using the
FP. Finally, in section~\ref{sec:sum}, we summarize our main results. We
use cosmological parameters $\Omega_0=0.27$, $\lambda=0.73$ and $h=0.7$
throughout the paper.

\section{The FP of galaxy clusters}
\label{sec:FP}

In Paper~I, we studied 20 massive clusters from the Cluster Lensing And
Supernova survey with Hubble (CLASH) observational dataset
\citep{pos12a,don14a,men14a,ume16a}. For these clusters, $r_s$ and $M_s$
were derived by gravitational lensing, and the X-ray temperatures $T_X$
were obtained by {\it Chandra} observations. We found that the clusters
form a two-parameter family and thus they are distributed on a plane in
the space of ($\log r_s, \log M_s, \log T_X$), which is described by
$a\log r_s + b\log M_s + c\log T_X=\mathrm{const.}$, with
$a=0.76^{+0.03}_{-0.05}$, $b=-0.56^{+0.02}_{-0.02}$, and
$c=0.32^{+0.10}_{-0.09}$ (Table~\ref{tab:para}). The plane normal is
represented by $P_3=(a,b,c)$, and the dispersion around the plane is
$0.045^{+0.008}_{-0.007}$ dex. The plane normal and the errors are
obtained through a principal component analysis (PCA) and Monte-Carlo
realizations.  From now on, we assume that the length of the plane
normal is $|P_3|=\sqrt{a^2 + b^2 + c^2}=1$ unless noted
otherwise. We showed in Paper~I that numerical simulations reproduce the
plane regardless of relaxation of clusters, redshifts, and gas physics
such as radiative cooling and active galactic nucleus (AGN) feedback. In
particular, clusters evolve within this plane and do not substantially
deviate from the plane even during major mergers.  This FP is consistent
with the one predicted from the similarity solution for structure
formation by \citet{ber85a}:
\begin{equation}
 \label{eq:FP}
 r_s^2 M_s^{-(n+11)/6}T_X={\rm const}\:,
\end{equation}
or $T_X\propto M_s^{(n+11)/6}/r_s^2$, where $n$ is the power spectrum
index of the initial density fluctuations of the universe (Paper~I).
The index is $n\sim -2$ at cluster scales \citep{eis98a,die15a}. Since
X-ray emissions mostly come from the inner region of clusters
($r\lesssim r_s$), $T_X$ mostly reflects the temperature therein. Thus,
it should depend on the gravitational potential that is represented by
$r_s$ and $M_s$, which is consistent with equation~(\ref{eq:FP}). Since
$r_s$ and $M_s$ are conserved in the slow-growing phase, it may be
natural to expect that $T_X$ is also conserved in that phase.

However, equation~(\ref{eq:FP}) is different from the one expected from
virial equilibrium at the formation of the inner structure or
$T_X\propto M_s/r_s$ (virial expectation). If a cluster is isolated and
steady, the virial theorem is given by $2K+W=0$, where $K$ is the
kinetic and/or thermal energy and $W$ is the gravitational energy.  This
should give the expression above ($T_X\propto M_s/r_s$). However, if
the cluster is not isolated and is growing, the virial theorem requires
a term of $d^2I/dt^2/2$, where $I$ is the moment of inertia, and two
boundary terms originating from the flux of inertia through the boundary
and the pressure at the boundary \citep{ber85a,shi16b}. The peculiar
relation of equation~(\ref{eq:FP}) is attributed to these effects. In
Paper~I, we also studied data distribution in the space of $(r_{200},
M_{200}, T_X)$, and confirmed that the data points follow the obvious
relation of $M_{200}\propto r_{200}^3$.

\section{Mass--temperature relation}
\label{sec:MT}

In this section, we derive the mass--temperature relation from the FP
relation and the concentration parameter $c_\Delta$.

The mass profile of a cluster is well described by the NFW formula and
can be derived from equation~(\ref{eq:NFW}):
\begin{equation}
\label{eq:MNFW}
 M(r) = 4\pi\delta_c\rho_c r_s^3
\left[\ln\left(1+\frac{r}{r_s}\right)-\frac{r}{r+r_s}\right]\:.
\end{equation}
From the definition of $M_\Delta$, we obtain
\begin{equation}
\label{eq:rD}
r_\Delta =
\left(\frac{3 M_\Delta}{4\pi \Delta\:\rho_c(z)}\right)^{1/3}\:.
\end{equation}
From equations~(\ref{eq:MNFW}) and~(\ref{eq:rD}), the normalization is
given by
\begin{equation}
\label{eq:delc}
 \delta_c = \Delta\: y(c_\Delta)\:,
\end{equation}
where
\begin{equation}
\label{eq:y}
 y(x) \equiv \frac{1}{3}\:\frac{x^3}{\ln(1+x)-x/(1+x)}\:,
\end{equation}
and the characteristic mass is given by
\begin{equation}
\label{eq:MDMs}
 M_s = M_\Delta\frac{\ln 2-1/2}{\ln(1+c_\Delta)-c_\Delta/(1+c_\Delta)}
\end{equation}
Equations~(\ref{eq:c}) and~(\ref{eq:MDMs}) show that the ratios
$r_s/r_\Delta$ and $M_s/M_\Delta$ are functions of $c_\Delta$.

Previous studies have shown that the concentration parameter $c_\Delta$
is a function of the mass $M_\Delta$ and the observed redshift $z$ of
a cluster. For example, \citet{duf08a} obtained an empirical relation
for $\Delta=200$ from $N$-body simulations:
\begin{equation}
\label{eq:duf08a}
 c_{200}(M_{200},z) = 6.71\:\left(\frac{M_{200}}
{2\times 10^{12}h^{-1}M_\odot}\right)^{-0.091}(1+z)^{-0.44}
\end{equation}
for $M_{200}\sim 10^{11}$--$10^{15}h^{-1}M_\odot$ and $z<2$ \citep[see
also][]{2013ApJ...766...32B,2014MNRAS.441.3359D,men14a,die15a}. \citet{cor15c}
considered the mass accretion history of dark halos and proposed an
analytical model based on the inside-out scenario.  Their model
reproduces equation~(\ref{eq:duf08a}) and is applicable even to smaller
$M_{200}$, larger $z$, and various cosmological parameters. We use
their code {\sf
COMMAH}\footnote{https://bitbucket.org/astroduff/commah}$^,$\footnote{Although
{\sf COMMAH} gives only $c_{200}$, it can be converted to $c_\Delta$ for
arbitrary $\Delta$ using the profile given by equation~(\ref{eq:NFW})}
to calculate $c_\Delta(M_\Delta,z)$.

\begin{figure*}
\epsscale{1.}\plottwo{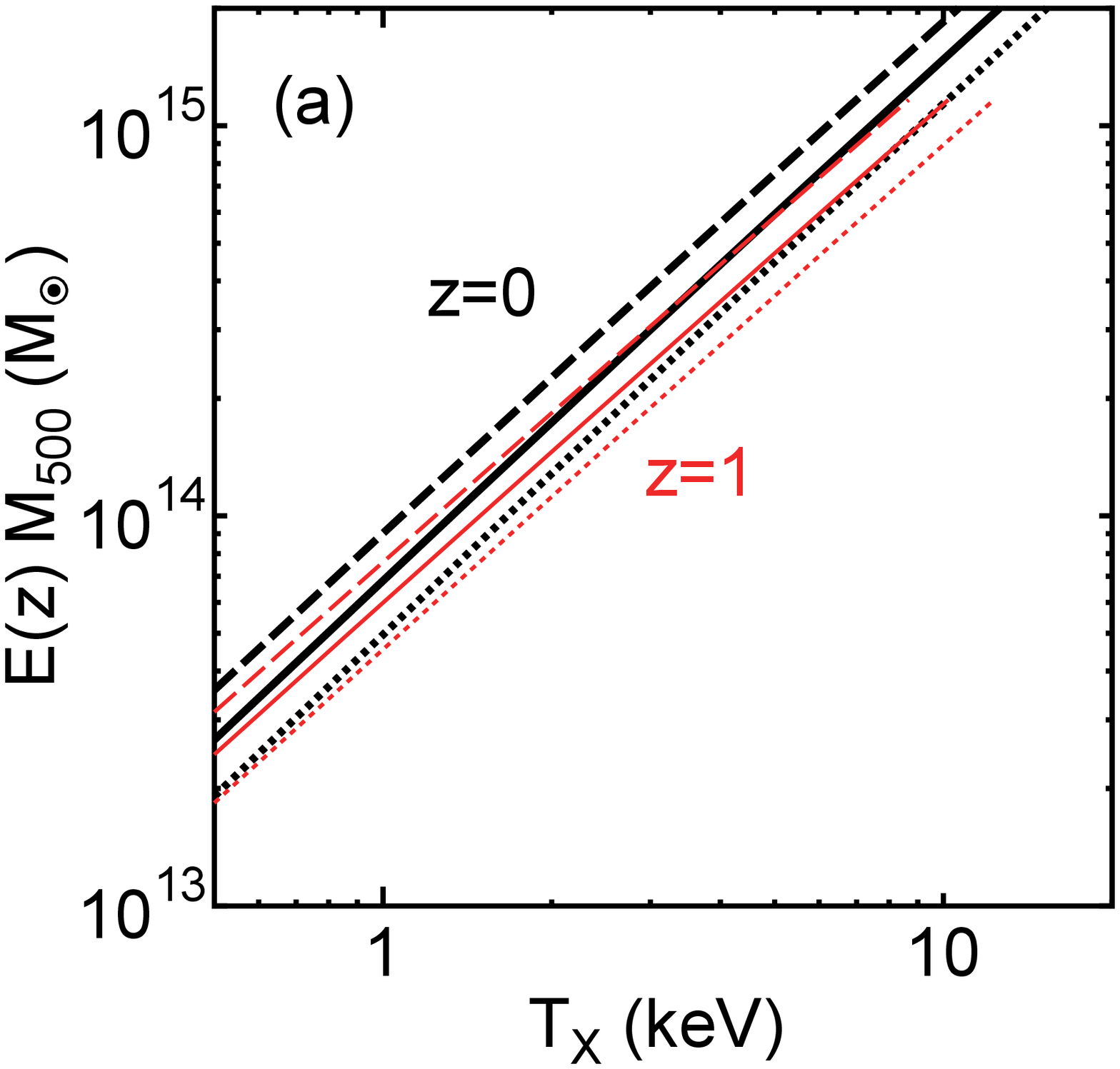}{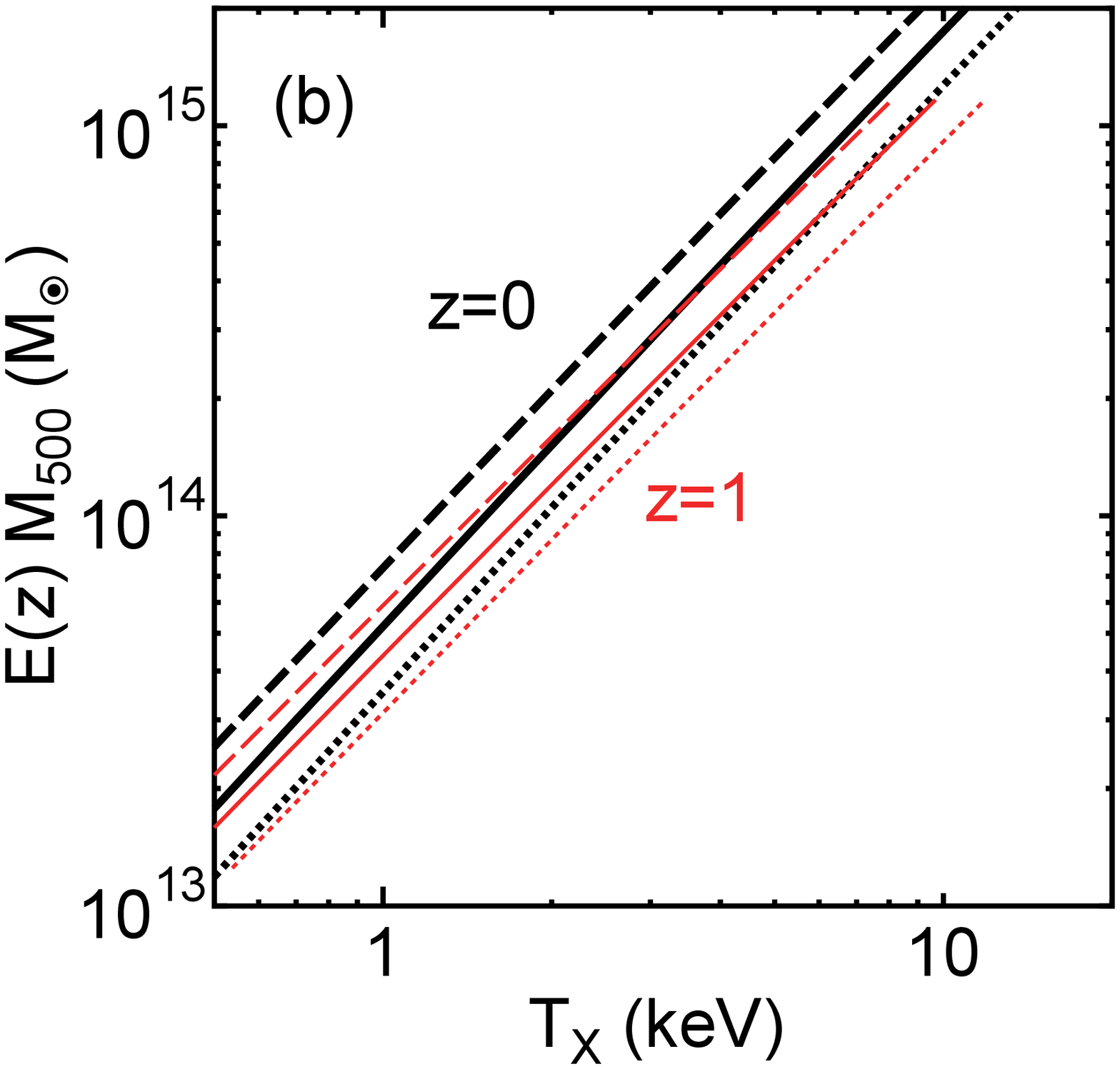}\\ \caption{(a) Relation
between $M_\Delta$ and $T_X$ for $\Delta=500$ and $n=-2$. The thick
black lines and the thin red lines represent $z=0$ and $z=1$,
respectively. The solid lines are calculated for the fiducial
$c_\Delta$. The dotted and dashed lines correspond to $c_\Delta^U$ and
$c_\Delta^L$, respectively. The slope of the relation at
$1\lesssim T_X\lesssim 7$~keV is $\alpha=1.33$ for $z=0$, and
$\alpha=1.28$ for $z=1$ ($E(z)M_{500}\propto T_X^\alpha$). (b) Same as
(a) but for $n=-2.5$. The slope of the relation at $1\lesssim
T_X\lesssim 7$~keV is $\alpha=1.53$ for $z=0$, and $\alpha=1.45$ for
$z=1$.\label{fig:MDTX}}
\end{figure*}

\begin{figure}
\epsscale{1.}\plotone{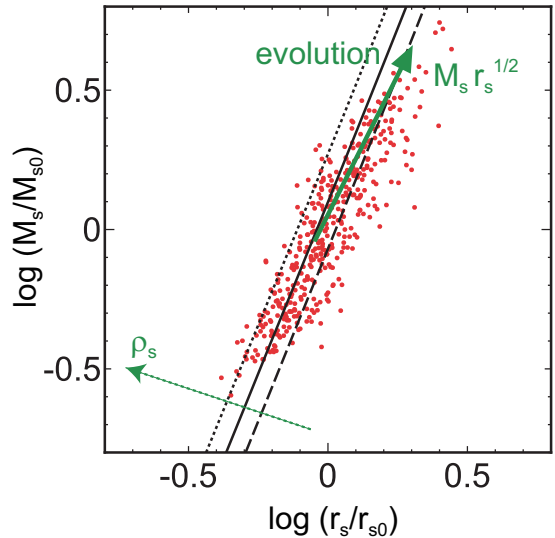} \caption{Relation between $M_s$ and
$r_s$ at $z=0.25$. The black solid line is calculated for the fiducial
$c_\Delta$. The black dotted and dashed lines are calculated for
$c_\Delta^U$ and $c_\Delta^L$, respectively.  Individual clusters
generally evolve in the direction of the green solid arrow to which the
value $M_s r_s^{1/2}$ increases. The characteristic density $\rho_s$
increases and the cluster formation epoch occurs earlier in the
direction of the green dotted arrow. Note that the dotted green
arrow is almost (but not perfectly) perpendicular to the three parallel
black lines. Red dots are the data points of the MUSIC simulation at
$z=0.25$, which are the same as those in Figure~5(a) of
Paper~I. \label{fig:rsMs}}
\end{figure}

For a given $z$ and $\Delta$, the $M_\Delta$--$T_X$ relation can be
obtained as follows. The mass $M_\Delta$ is converted to $M_s$ by
equation~(\ref{eq:MDMs}) and $c_\Delta(M_\Delta,z)$. The radius
$r_\Delta$ is a function of $M_\Delta$ (equation~(\ref{eq:rD})), and
it is converted to $r_s$ by equation~(\ref{eq:c}) and
$c_\Delta(M_\Delta,z)$. Here, we use the analytically derived FP
(equation~(\ref{eq:FP})). Thus, the temperature is given by
\begin{equation}
\label{eq:TX}
 T_X = T_{X0}\left(\frac{r_s}{r_{s0}}\right)^{-2}
\left(\frac{M_s}{M_{s0}}\right)^{(n+11)/6}\:,
\end{equation}
where $(r_{s0}, M_{s0}, T_{X0})$ correspond to a representative point on
the FP. Since the similarity solution does not predict $(r_{s0}, M_{s0},
T_{X0})$, we adopt the logarithmic mean of the parameters for
the MUSIC simulation sample; $r_{s0}=414$~kpc, $M_{s0}=1.4\times
10^{14}\: M_\odot$, and $T_{X0}=3.7$~keV (Table~\ref{tab:para}, see also
Figure~\ref{fig:rsMs}). The MUSIC simulation set that we consider is
from non-radiative runs and includes 402 clusters at $z=0.25$ with
$M_{200}>2\times 10^{14}\: h^{-1}\: M_\odot$ (Paper~I, see also
\citealp{men14a}). Assigning $r_s$ and $M_s$ in equation~(\ref{eq:TX}),
we finally obtain the $M_\Delta$--$T_X$ relation. Numerical simulations
have shown that the $c_\Delta$--$M_\Delta$ relation has an intrinsic
scatter of $\sim 0.1$~dex at $M_{200}\sim 10^{14}$--$M^{15}\:\rm
M_\odot$ \citep{bul01a,duf08a,lud13a,men14a,cor15c}. Thus, we also
calculate the $M_\Delta$--$T_X$ relations when $c_\Delta$ (fiducial) is
replaced by $c_\Delta^U = 10^{0.1} c_\Delta$ (upper limit) or
$c_\Delta^L = 10^{-0.1} c_\Delta$ (lower limit). In the cluster
mass range the scatter of the relation is not particularly sensitive to
the baryonic physics or to the fitting radial range. In fact,
\citet{ras13a} found an intrinsic scatter of $\sim$0.1 dex for the
relations obtained from simulations both with and without AGN feedback
(Tables~1 and~2 of \citealt{ras13a}). Figure~\ref{fig:MDTX} shows the
relation $M_\Delta$--$T_\Delta$ for $\Delta=500$ at $z=0$ and~1. We
adopt $n=-2$ for Figure~\ref{fig:MDTX}(a). The slope of the
lines at $1\lesssim T_X\lesssim 7$~keV is $\alpha=1.33$ for $z=0$
($E(z)M_{500}\propto T_X^\alpha$). The values are close to $3/2$ or
$1.5$, and $c_\Delta^U$ gives smaller $M_{500}$ than $c_\Delta^L$. This
relation holds even when $\Delta=200$ and 2500. We emphasize that we did
not use the assumption of virial equilibrium when we derive the
relation. The figure also shows that the vertical dispersion of the
relations should be within a factor of two. The slight difference of the
obtained slope $\alpha$ from $1.5$ may be due to some simplified
assumption we made when we derived the plane angle. For example,
equation~(\ref{eq:FP}) is exactly correct only for the Einstein–de
Sitter universe because it is based on a similarity solution. For the
$\Lambda$CDM cosmology we adopted, the angle of the plane could slightly
change (\citealp{ber85a}; Paper~I, see also \citealp{shi16a}).

In fact, the slope of $\alpha\sim 1.5$ can be obtained if we
change the plane angle only slightly. Figure~\ref{fig:MDTX}(b) is the
same as Figure~\ref{fig:MDTX}(a) but for $n=-2.5$ in
equation~(\ref{eq:FP}). While the plane angle is almost the same as that
for $n=-2$ (Table~\ref{tab:para}) and is consistent with the
observations (see Figure~\ref{fig:prob}), the slope of the lines is
$\alpha=1.53$ for $z=0$. Thus, an imperceptible modification of the
angle is enough to obtain $\alpha\sim 1.5$. We also note that the power
spectrum index $n$ is expected to be smaller at smaller scales. For
example, $n\sim -2.5$ is expected at group scales ($M_{200}\sim
10^{13}\: M_\odot$; e.g. \citealt{die15a}). This means that the
$M_\Delta$--$T_X$ relation should become qualitatively steeper toward
lower $T_X$.

Figure~\ref{fig:rsMs} shows the relation between $r_s$ and $M_s$ at
$z=0.25$; they are calculated using $c_{200}(M_{200},z)$. We also
plot the data points of the MUSIC simulation (Paper~I). For the
simulation, the scale radius $r_s$ is obtained by fitting the total
density distribution (gas+dark matter) with the NFW profile up to
$r_{200}$. The mass $M_s$ is then derived as the mass enclosed by a
sphere of radius $r_s$. The figure is similar to Figure~5(a) of Paper~I
and it is the projection of the FP on the $r_s$--$M_s$ plane. As can be
seen, most of the data points are distributed inside the upper and lower
limits of $c_\Delta$. This clearly indicates that the dispersion of the
$c_\Delta$--$M_\Delta$ relation corresponds to the spread of the cluster
distribution along the minor axis of the FP. Note that the limits along
the major axis or the larger and smaller $M_s$ limits of the MUSIC data
distribution are set by the box size and the resolution of the
simulation, respectively. The characteristic density $\rho_s$ increases
in the direction of the dotted green arrow. Since the three black lines
are almost (but not perfectly) perpendicular to the arrow
representing the direction of the $\rho_s$ axis, the variation of
$c_\Delta$ is closely related to that of $\rho_s$ or the formation time
of clusters, although it is not a precise one-to-one
correspondence. Individual clusters evolve toward lower $\rho_s$ as a
whole, but the actual direction (solid green arrow) is mostly determined
by the power spectrum of the initial density fluctuations of the
universe, and cluster mergers temporally disturb this motion on the FP
(Paper~I).

\begin{figure*}
\plottwo{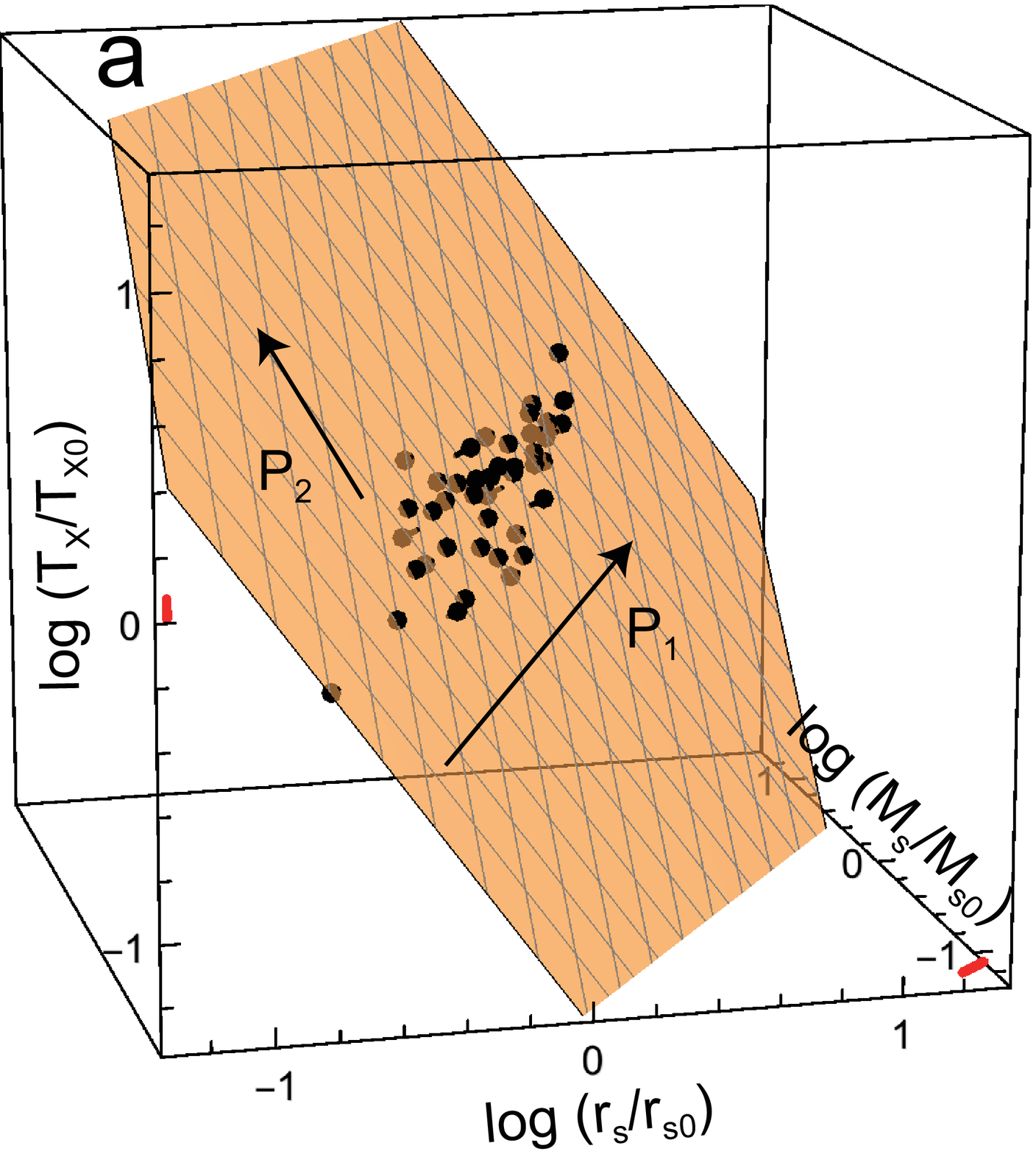}{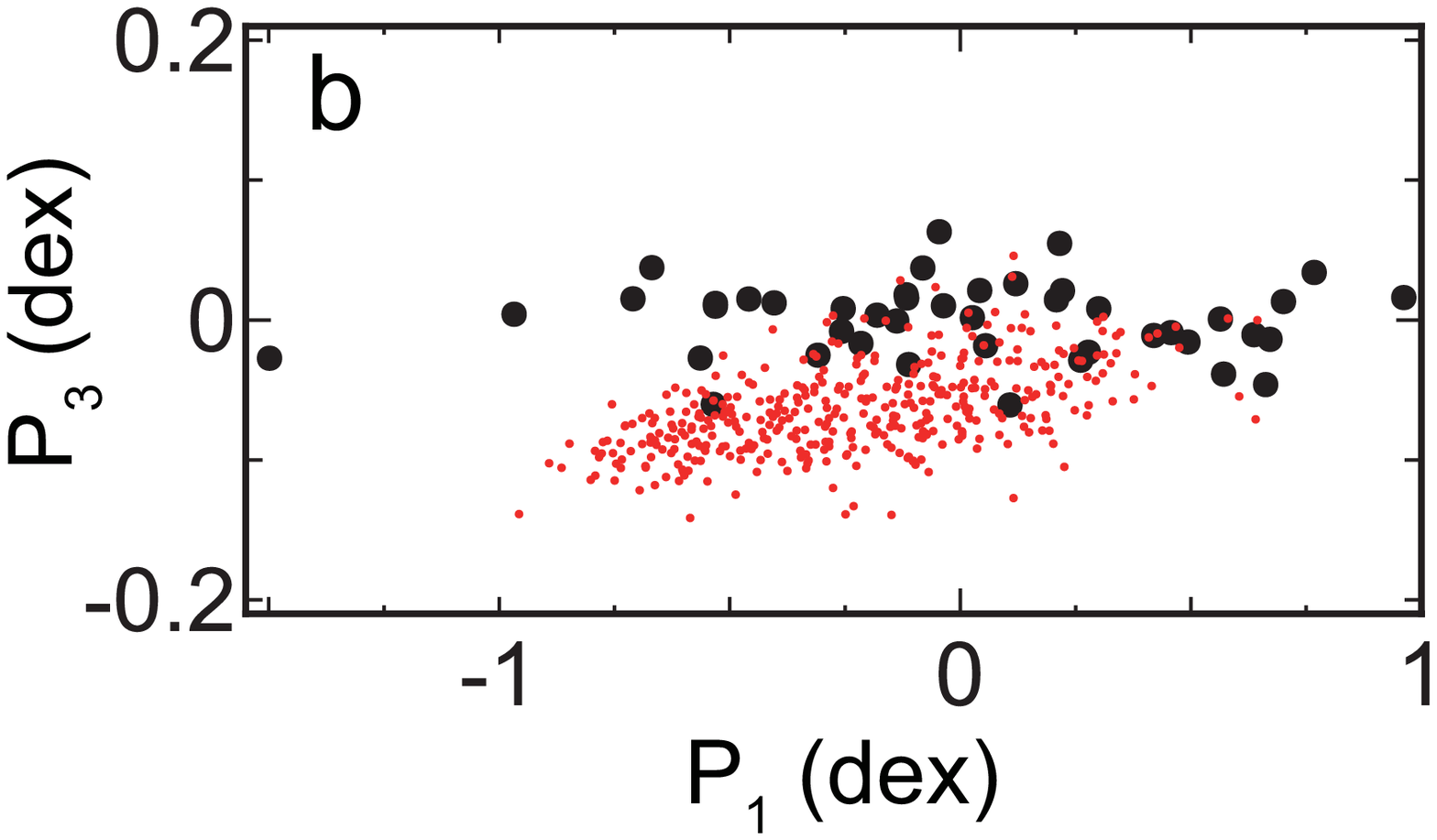} \caption{(a) Points (pin heads) show the
distribution of the 44 X-ray clusters in the space of $(\log
(r_s/r_{s0}), \log (M_s/M_{s0}), \log (T_X/T_{X0}))$, where
$r_{s0}=443$~kpc, $M_{s0}=2.2\times 10^{14}\: M_\odot$, and
$T_{X0}=7.3$~keV are the sample geometric averages (log means) of $r_s$,
$M_s$, and $T_X$, respectively. The length of a pin shows the distance
between the point and the obtained plane. The orange plane is
translucent and grayish points are located below the plane. The arrow
$P_1$ shows the direction on the plane in which the data are most
extended, and the arrow $P_2$ is perpendicular to $P_1$ on the
plane. The red bars at the corner of the $\log r_s$--$\log M_s$ plane
and on the $\log T_X$ axis are typical $1\sigma$ errors of the data. (b)
Cross-section of the plane shown in (a). The origin is the same as (a)
and $P_3$ is the plane normal. The large black points are the X-ray
clusters shown in (a). The small red points are the MUSIC-simulated
clusters projected on the $P_1$--$P_3$ plane determined for the X-ray
clusters.\label{fig:plane}}
\end{figure*}

\begin{figure}
\epsscale{1.}\plotone{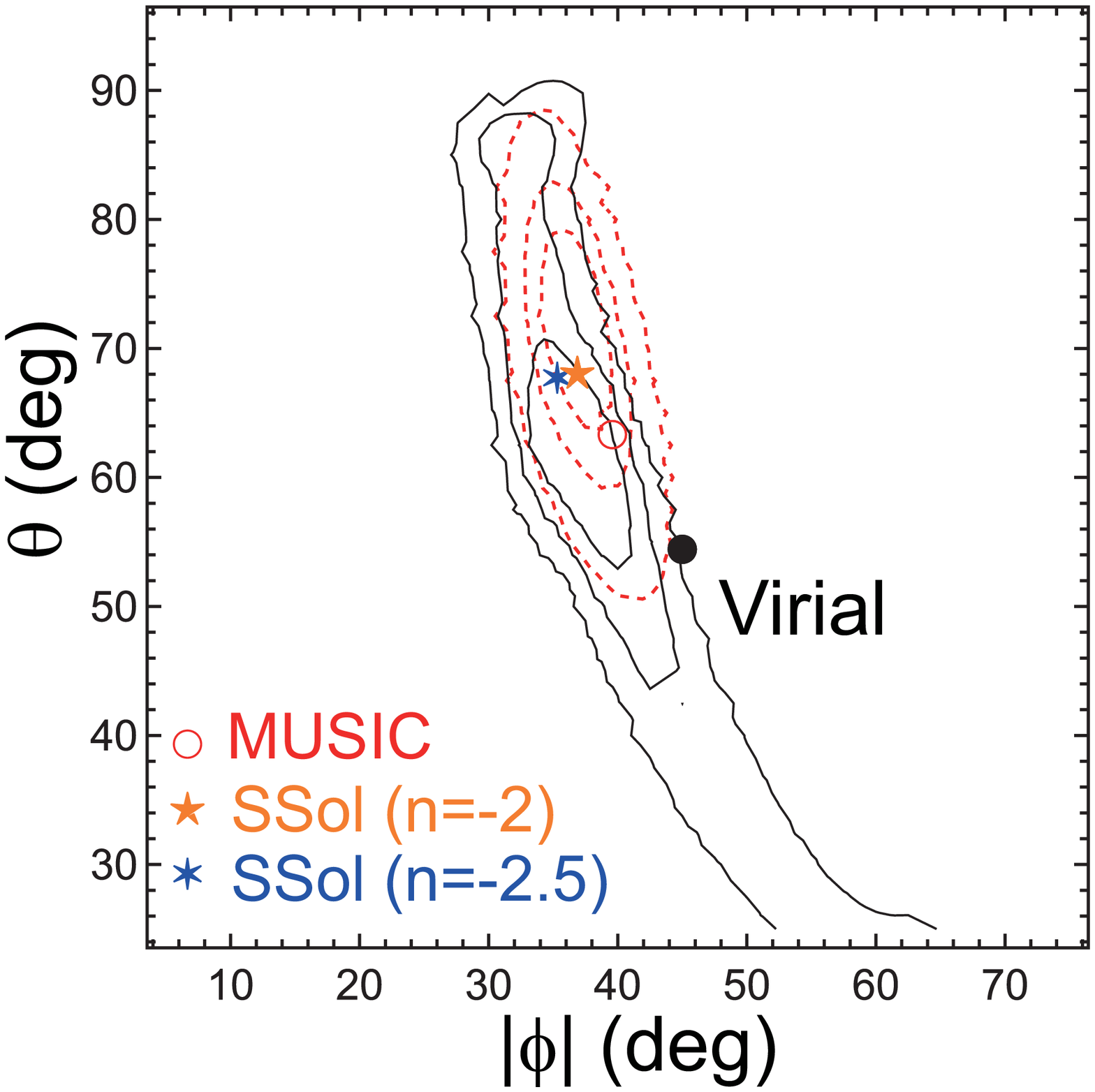} \caption{The direction of the plane
normal $P_3=(a,b,c)$ in the space of $(\log r_s, \log M_s, \log T_X)$.
Probability contours for the sample of 44 X-ray clusters (black solid
lines) are shown at the 68 ($1\:\sigma$), 90, and 99\% confidence levels
from inside to outside. The prediction of the virial expectation ($r_s
M_s^{-1} T_X \propto \rm const$) corresponds to $(\phi,
\theta)=(-45^\circ, 55^\circ)$ (black dot), and is rejected at the 99\%
confidence level. Probability contours for the CLASH sample (red dotted
lines; see Figure~2 of Paper~I) are shown for comparison. The plane
normal derived for the MUSIC simulation sample is shown by the open red
circle; it is located around the 68\% contour level and consistent with
the X-ray observations at that level. Predictions based on a
similarity solution (SSol) for $n=-2$ and $n=-2.5$ are shown by the
orange and the blue star, respectively.\label{fig:prob}}
\end{figure}

\section{FP in X-rays}
\label{sec:XFP}

Since X-ray data have shown that the relation $M_\Delta\propto
T_X^{3/2}$ is generally satisfied
\citep{1999MNRAS.305..834E,1999A&A...348..711N,2000ApJ...532..694N,2001A&A...368..749F,2001ApJ...553...78X,2006ApJ...640..691V},
we expect that X-ray data alone form the FP. We will confirm this in
this section.

The characteristic radius $r_s$ and mass $M_s$ can be derived from X-ray
data, assuming that the ICM is in hydrostatic equilibrium. Using the
X-ray data of 44 clusters obtained with {\it XMM-Newton}
(\citealt{ett10a}; Table~\ref{tab:data}), we study their distribution in
the space of ($\log r_s, \log M_s, \log T_X$). The average redshift of
the clusters is 0.189. The logarithmic means of $(r_s, M_s, T_X)$ for
this sample are $r_{s0}=443$~kpc, $M_{s0}=2.2\times 10^{14}\: M_\odot$,
and $T_{X0}=7.3$~keV (Table~\ref{tab:para}). While we adopt the values
obtained through method~1 in \citet{ett10a}, the results are not much
different even if we use those obtained through method~2. In
Table~\ref{tab:data}, the X-ray temperatures ($T_X$) and masses
($M_{200}$) obtained by \citet{ett10a} are listed as $T_{\rm XMM}$ and
$M_{\rm 200,XMM}$, respectively. The temperatures are the error-weighted
mean of the spectral measurements in the radial range $[0.15\: r_{500},
{\rm min}(r_{500},R_{\rm xsp})]$, where $R_{\rm xsp}$ is the maximum
radius up to which X-ray spectra can be extracted (see Table~2 of
\citealt{ett10a}). This means that $M_{\rm 200,XMM}$ and $c_{200}$ are
estimated from the X-ray emission at $\leq r_{500}$. It has been
indicated that the temperatures obtained with {\it XMM-Newton} tend to
be lower than those obtained with {\it Chandra}
\citep{2010A&A...523A..22N,don14a,2015MNRAS.448..814I,2015A&A...575A..30S,zha15b}. Since
the temperatures we used for the CLASH sample have been derived with
{\it Chandra} \citep{pos12a}, a correction is required to compare the
temperatures between the two samples. \citet{ett10a} used only
MOS1+MOS2, with MOS2 as a value of reference. We convert $T_{\rm XMM}$
in Table~\ref{tab:data} into the equivalent {\it Chandra} temperature
$T_{\rm Ch}$ using the relation,
\begin{equation}
\label{eq:Tch}
 \log\frac{T_{\rm Ch}}{\rm keV} 
= \frac{1}{A}\left(\log\frac{T_{\rm XMM}}{\rm keV}-B\right)\;,
\end{equation}
where $A=0.909^{+0.005}_{-0.005}$ and $B=-0.017^{+0.003}_{-0.004}$
\citep{2015A&A...575A..30S}.  Hereafter, we consider $T_X$ as the
corrected temperature, $T_{\rm Ch}$. Moreover, since cluster mass is
estimated based on the temperature, we convert $M_{\rm 200,XMM}$ in
Table~\ref{tab:data} into the equivalent {\it Chandra} mass $M_{\rm
200,Ch}$ using the relation of $M_{\rm 200,Ch}=(T_{\rm Ch}/T_{\rm
XMM})M_{\rm 200,XMM}$ and we refer to $M_{\rm 200,Ch}$ as $M_{200}$.
The correction does not really affect the following results because we
discuss the FP in the logarithmic space. We calculate $M_s$ from
$M_{200}$ and $c_{200}$ using equation~(\ref{eq:MDMs}) assuming that
$c_{200}$ is anti-correlated with $r_s$ (Figure~2 of \citealt{ett10a}).

In Figure~\ref{fig:plane}(a), we show the results for the whole sample
of 44 clusters; the data points have a planar distribution. The cross
section of the plane is shown in Figure~\ref{fig:plane}(b). In this
figure, we project the simulated MUSIC clusters on the cross section of
the X-ray cluster plane for comparison. The distribution of the MUSIC
clusters is slightly deviated from that of the X-ray clusters. We will
discuss the absolute position of the X-ray cluster plane in
Section~\ref{sec:calib}.  We determine the direction of the X-ray
cluster plane and the errors through a PCA and Monte-Carlo realizations
(see Paper~I). The direction on the plane in which the data are most
extended is $P_1$, and the direction perpendicular to $P_1$ on the plane
is $P_2$. The plane is described by $a\log r_s + b\log M_s + c\log
T_X=\mathrm{const.}$, with $a=0.71^{+0.06}_{-0.10}$,
$b=-0.53^{+0.02}_{-0.01}$, and $c=0.46^{+0.13}_{-0.11}$. The values of
$(a,b,c)$ are consistent with those for the CLASH sample within the
errors (Table~\ref{tab:para}). The thickness of the plane or the
dispersion in the direction of $P_3$ is $0.039^{+0.021}_{-0.010}$, which
is slightly smaller than, but consistent within the errors with the
CLASH result (Paper~I). Here we note that lensing measurements are
sensitive to projection effects, and NFW fitting based on the assumption
of spherical symmetry can introduce a sizable scatter in the derived
mass and concentration parameters, or $(r_s, M_s)$. In our error
analysis of the CLASH lensing data \citep{ume16a}, we properly accounted
for the projection effects due to cluster halo triaxiality and
uncorrelated large-scale structure projected along the line of sight
\citep{2015MNRAS.449.4264G}, implying that the thickness of the FP
derived from our CLASH data should not be affected by the external
scatter introduced by the lensing projection effects. The X-ray sample
has a wider range of $r_s$ and $M_s$ compared with the CLASH sample
(Figure~1 in Paper~I). Since the errors of $a$, $b$, and $c$ are not
independent of each other, we show in Figure~\ref{fig:prob} the
likelihood contours of the parameters describing the direction of the
plane normal $P_3$ for the X-ray sample (black solid lines). In that
figure, $\theta$ is the angle between $P_3$ and the $\log T_X$ axis, and
$\phi$ is the azimuthal angle around the $\log T_X$ axis, measured
anti-clockwise from the $\log r_s$ axis, or $\tan\phi=b/a$
(Table~\ref{tab:para}). The contours are elongated in the direction of
rotation around $P_1$ (Figure~\ref{fig:plane}(a)), to which the
direction $P_3$ is less constrained. The contours show that the
direction of $P_3$ is consistent with that for the CLASH sample (red
dotted lines; see Paper~I). As long as clusters are widely distributed
on the plane, the direction of $P_3$ should not be too affected by a
possible sample selection bias. In Figure~\ref{fig:prob}, we also
plotted MUSIC simulation results (see Paper~I for details) and the
prediction of the similarity solution (SSol), which is given by
equation~(\ref{eq:FP}) with $n=-2$ and $-2.5$. They are the same as
those in Figure~2 of Paper~I and are consistent with the X-ray data at
the $\sim 1\:\sigma$ level. For the virial expectation, the angle
$\theta$ is the one between vectors
$(1/\sqrt{3},-1/\sqrt{3},1/\sqrt{3})$ and $(0,0,1)$, which is $\approx
55^\circ$. The prediction of the virial expectation is rejected at the
99\% confidence level.

In Paper~I, using the results of numerical simulation, we showed that
the plane parameters are not very dependent on the relaxation state of
clusters, although irregular clusters tend to slightly increase the
scatter of the FP. That is, although the cluster parameters ($r_s$,
$M_s$, $T_X$) can fluctuate substantially, especially during major
mergers, the particular combination of these parameters that determines
the FP (e.g. the left side of equation~\ref{eq:FP}) can remain nearly
constant. As a result, clusters evolve along the FP and do not much
deviate from the FP even during a cluster merger (Section~5.2 in
Paper~I). Here, we study this issue using the X-ray
sample. \citet{ett10a} classified the X-ray sample into 18 cool-core
(CC) clusters, 19 non-cool-core (NCC) clusters, and 7 intermediate
cool-core (ICC) clusters based on the entropy of the ICM in the central
region. The CC and NCC clusters tend to be regular and irregular in
shape, respectively. We performed the FP analysis for the CC and ICC+NCC
samples separately and the results are shown in
Table~\ref{tab:para}. The plane directions are not much different from
that of the whole sample of 44 clusters. The thickness of the plane is
$0.026^{+0.014}_{-0.006}$ for CC and $0.043^{+0.019}_{-0.010}$ for
ICC+NCC, which can be compared with the one for the whole sample
($0.039^{+0.021}_{-0.010}$). Since $r_s$ is generally much larger than
the cluster core, the details of the ICM physics in the core region are
not expected to significantly affect the global cluster parameters,
$(\log r_s, \log M_s, \log T_X)$. Although the thicknesses of the planes
for the different subsamples are consistent with each other within
errors, CC (ICC+NCC) clusters may give a smaller (larger) dispersion
about the plane. We note that while the size of the ICC sample alone is
too small to reliably determine their FP parameters, the thickness of
the plane is slightly increased by including the ICC sample compared to
the NCC-only case.

\section{Shift of the planes and mass calibration}
\label{sec:calib}

\begin{figure}
\epsscale{1.}\plotone{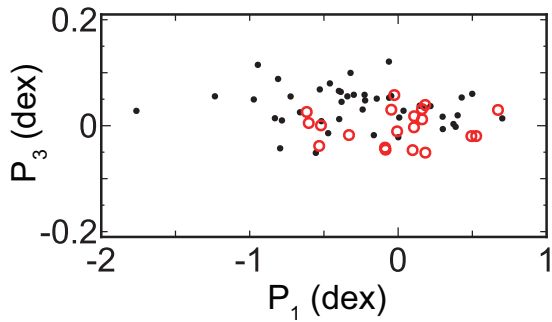} \caption{The cross-section of the CFP
(red circles) and the XFP (black dots). The coordinates $P_1$ and
$P_3$ are for the CFP. \label{fig:P13_mix}}
\end{figure}

Figure~\ref{fig:P13_mix} shows the cross sections of the FPs for the
CLASH and X-ray samples depicted on the same plane coordinate. Although
they overlap with each other, the FP for the CLASH sample (CFP) is
located slightly above the FP for the X-ray sample (XFP). Fixing the
angles of both planes to the same one given by equation~(\ref{eq:FP})
with $n=-2$ (SSol in Table~\ref{tab:para}), we estimate the distance
between the two FPs and find that it is $d_{\rm
FP}=0.031^{+0.027}_{-0.039}$~dex in the space of $(\log r_s, \log M_s,
\log T_X)$. Thus, the shift of the two planes is not significant. The
error of the distance mostly comes from the observational errors of the
X-ray data, which result in the uncertainty of the position of the XFP.

However, X-ray data will be enriched and the configuration of
the FP could be determined much more precisely in the near future. In
principle, the FP can be used as a benchmark of data calibration,
because numerical simulations have shown that the FP is very thin and
its origin has been explained by the similarity solution (Paper~I).
Here, we demonstrate that a possible shift of the FP could be used to
calibrate cluster masses $M_\Delta$ obtained with X-ray observations. It
would be useful even for studies of cluster number counts based on
masses estimated by the Sunyaev-Zel'dovich effect, because this
estimation relies on X-ray data for calibration \citep{pla14b}. In the
following, we use the current observational datasets, although they may
not be accurate enough for our calibration purposes.

We study the shift of the XFP against the CFP assuming that the shift is
caused by some observational systematic errors. Given the fairly large
statistical uncertainties, and for the sake of simplicity, the angles of
both FPs are fixed (SSol for $n=-2$ in Table~\ref{tab:para}), because
the direction is consistent with both the CFP and XFP
(Figure~\ref{fig:prob}) and the distance between two planes can be well
defined only when the planes are parallel. If the data quality and size
are improved in the future, this constraint may not be needed. Since
both FPs use X-ray data for the temperature, and since the temperature
has been corrected by equation~(\ref{eq:Tch}), we assume that there is
no systematic error for temperature, for simplicity. In that case, the
shift of the planes may be attributed to the systematic error of $M_s$
and/or $r_s$. The errors of $M_s$ and $r_s$ may come from the assumption
of hydrostatic equilibrium and the limited radial range adopted in X-ray
analysis, respectively \citep[e.g.][]{2007ApJ...655...98N,ras13a}.
First, let us assume that the XFP is shifted solely in the direction of
$M_s$. In this case, the positions of a given cluster on the CFP and the
XFP are given by $(r_s, M_{sC}, T_X)$ and $(r_s, M_{sX}, T_X)$,
respectively. From now on, we shall use subscript $C$ or $X$ if the
value is specifically related to the CFP or the XFP, respectively. Since
the two FPs are parallel, the ratio $f_{Ms}\equiv M_{sX}/M_{sC}$ is not
unity but is independent of clusters. However, the ratio of $M_\Delta$
among the two FPs or $f_{M\Delta}\equiv M_{\Delta X}/M_{\Delta C}$ can
vary because $M_\Delta$ is a function of $c_\Delta$
(equation~(\ref{eq:MDMs})) that is not constant among clusters. In the
Appendix~\ref{sec:app}, we show that $f_{M\Delta}$ is represented by a
function of $c_{\Delta X}$ (equation~\ref{eq:fMD-cDX}) or $c_{\Delta C}$
(equation~\ref{eq:fMD-cDC}) for a given $f_{Ms}$. Second, let us assume
that the XFP is shifted solely in the direction of $r_s$. In this case,
the positions of a given cluster on the CFP and the XFP are given by
$(r_{sC}, M_s, T_X)$ and $(r_{sX}, M_s, T_X)$, respectively. The ratio
$f_{rs}\equiv r_{sX}/r_{sC}$ is not unity but is independent of
clusters. For the second case, we can also derive $f_{M\Delta}$ as a
function of $c_{\Delta X}$ (equation~\ref{eq:fMD2}) or $c_{\Delta C}$
(equation~\ref{eq:fMD-cDC2}) for a given $f_{rs}$.

\begin{figure}
\epsscale{1.}\plotone{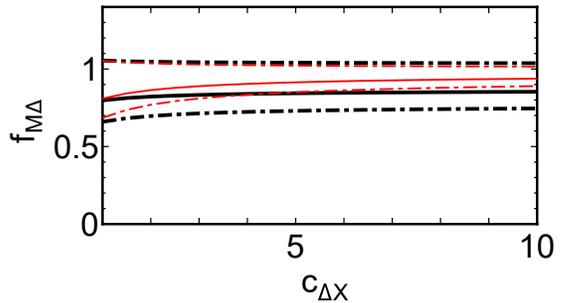} \caption{Relation between $f_{M\Delta}$
and $c_{\Delta X}$. The thick black lines are derived from
equation~(\ref{eq:fMD-cDX}) and thin red lines are derived from
equation~(\ref{eq:fMD2}). The solid lines are for the most certain
values of $f_{Ms}$ (thick black) or $f_{rs}$ (thin red). The
dashed-dotted lines show the uncertainties attributed to those of
$f_{Ms}$ (thick black) or $f_{rs}$ (thin red). \label{fig:McalibX}}
\end{figure}

Figure~\ref{fig:McalibX} shows the relation between $f_{M\Delta}$ and
$c_{\Delta X}$ for our CLASH and X-ray samples. The results do not
depend on the value of $\Delta$. Since we assumed that the normal of the
two FPs is given by equation~(\ref{eq:FP}) for $n=-2$, it is written as
${P_3}=(a,b,c)=(0.74, -0.56, 0.37)$ (SSol in Table~\ref{tab:para}). If
the shift of the FP is caused by a systematic error of $M_s$, we have
$f_{Ms}=10^{d_{\rm FP}/b}=0.88^{+0.15}_{-0.09}$. Thus, $f_{M\Delta}$ can
be derived from equation~(\ref{eq:fMD-cDX}), which is shown by the thick
black lines in Figure~\ref{fig:McalibX}. On the other hand, if the shift
of the FP is caused by a systematic error of $r_s$, we have
$f_{rs}=10^{d_{\rm FP}/a}=1.10^{+0.10}_{-0.12}$. Thus, $f_{M\Delta}$ is
derived from equation~(\ref{eq:fMD2}), which is shown by the thin red
lines in Figure~\ref{fig:McalibX}. The $f_{Ms}$--$c_{\Delta C}$ relation
is almost the same as the $f_{Ms}$--$c_{\Delta X}$ relation.

Figure~\ref{fig:McalibX} shows that $f_{M\Delta}$ does not much depend
on $c_{\Delta X}$ and the dependence can be ignored, given the accuracy
of the current observations. The dashed-dotted lines suggest that the
uncertainty caused by the error of $f_{Ms}$ (black dashed-dotted lines)
is larger than that caused by the error of $f_{rs}$ (red dashed-dotted
lines). The actual uncertainty should be between the two. If we
conservatively adopt the former, $f_{M\Delta}=M_{\Delta X}/M_{\Delta
C}\sim 0.85_{-0.2}^{+0.2}$, which means that the mass estimated from the
X-ray data may be systematically underestimated compared with that
estimated from the CLASH data, but the evidence is not solid. The value
of $f_{M\Delta}$ we obtained is consistent with that predicted by
numerical simulations
\citep{2007ApJ...655...98N,2008A&A...491...71P,2010A&A...510A..76L,2012NJPh...14e5018R}. Future
improvements of observational data, especially X-ray data, are desired
to make a firm conclusion. For the analysis of those data, the
dependence of $f_{M\Delta}$ on $c_\Delta$ will need to be taken into
account. The direction of the plane shift could be determined if $r_s$
and $M_s$ of certain clusters were precisely determined from both X-ray
and gravitational lensing observations. Mock observations of simulated
clusters would also be useful to study the shift. Possible differences
in the angles of the CFP and the XFP could be a clue for identifying the
origin of the observational systematic errors. It would also be
interesting to compare the XFPs obtained with different instruments
(e.g. {\it Chandra} and {\it XMM-Newton}) for calibration including
$T_X$.

\section{Summary}
\label{sec:sum}

We have investigated the origin of the mass--temperature relation of
galaxy clusters. Observations and numerical simulations have shown that
the relation is approximately represented by $M_\Delta \propto
T_X^{3/2}$ (e.g. $\Delta=500$). This relation has been interpreted as
evidence that clusters are in virial equilibrium within
$r_\Delta$. However, the existence of the FP of clusters and its
interpretation based on the modern inside-out scenario suggest that
clusters are not in virial equilibrium in the whole region within
$r_\Delta$ and that the temperatures of clusters are primarily
determined by the characteristic mass $M_s$ and radius $r_s$ of the NFW
profile rather than $M_\Delta$. We have solved this discrepancy by
combining the FP with the concentration--mass--redshift relation of
cluster halos calibrated by $N$-body simulations. The relation $M_\Delta
\propto T_X^{3/2}$ is derived from the FP relation among $r_s$, $M_s$,
and $T_X$ using the mass dependence of $c_\Delta$. We also showed that
the dispersion of the $c_\Delta$--$M_\Delta$ relation can largely
account for the spread of the cluster distribution on the FP. Moreover,
we confirmed that the FP constructed from X-ray data alone is consistent
with that from gravitational lensing data. The FP could be used to
calibrate the cluster parameters derived with different methods. As an
example, we demonstrated that a cluster mass derived from X-ray
observations is systematically $\sim 85_{-20}^{+20}$~\% of that derived
from gravitational lensing observations.

\acknowledgments

We thank the anonymous referee whose comments greatly improved the
clarity of this paper. This work was supported by MEXT KAKENHI
No.~15K05080 (Y.F.). K.U. acknowledges support from the Ministry of
Science and Technology of Taiwan (grant MOST 106-2628-M-001-003-MY3) and
from the Academia Sinica Investigator Award. S.E. acknowledges financial
contribution from the contracts NARO15 ASI-INAF I/037/12/0, ASI
2015-046-R.0 and ASI-INAF n.2017-14-H.0. E.R. acknowledges support from
the ExaNeSt and EuroExa projects, funded by the European Union’s
Horizon 2020 research and innovation program under grant agreements No
671553 and No 754337, respectively.


\begin{deluxetable*}{ccccccccc}[ht]
\tablecaption{Plane parameters\label{tab:para}}
\tablenum{1}
\tablewidth{0pt}
\tablehead{
\colhead{Sample} &
\colhead{$a$} & \colhead{$b$} & \colhead{$c$} & 
\colhead{$\phi$} & \colhead{$\theta$} &
\colhead{$r_{s0}$} & \colhead{$M_{s0}$} & \colhead{$T_{X0}$}\\
\colhead{} &
\colhead{} & \colhead{} & \colhead{} & 
\colhead{(degree)} & \colhead{(degree)} &
\colhead{(kpc)} & \colhead{($10^{14}\: M_\odot$)} & \colhead{(keV)}
}
\startdata
CLASH (CFP) & 
$0.76^{+0.03}_{-0.05}$ & $-0.56^{+0.02}_{-0.02}$
& $0.32^{+0.10}_{-0.09}$ & -37 & 71 & 570 & 3.8 & 8.2\\
SSol\tablenotemark{a} ($n=-2$) &
0.74 & -0.56 & 0.37 & -37 & 68 & \nodata & \nodata & \nodata \\
SSol\tablenotemark{a} ($n=-2.5$) &
0.76 & -0.54 & 0.38 & -35 & 68 & \nodata & \nodata & \nodata \\
Virial &
0.58 & -0.58 & 0.58 & -45 & 55 & \nodata & \nodata & \nodata \\
MUSIC &
0.69 & -0.57 & 0.44 & -40 & 64 & 414 & 1.4 & 3.7 \\
X-ray (XFP)\tablenotemark{b} &
$0.71^{+0.06}_{-0.10}$ & $-0.53^{+0.02}_{-0.01}$ 
& $0.46^{+0.13}_{-0.11}$ & -37 & 63 & 443 & 2.2 & 7.3 \\
X-ray (CC)\tablenotemark{c} &
$0.72^{+0.04}_{-0.09}$ & $-0.52^{+0.02}_{-0.01}$ 
& $0.45^{+0.12}_{-0.05}$ & -36 & 63 & 529 & 2.7 & 7.1 \\
X-ray (ICC+NCC)\tablenotemark{d} &
$0.69^{+0.10}_{-0.07}$ & $-0.53^{+0.03}_{-0.02}$ 
& $0.49^{+0.10}_{-0.23}$ & -37 & 61 & 392 & 1.9 & 7.5 \\
\enddata
\tablenotetext{a}{Similarity solution (equation~(\ref{eq:FP})).} 
\tablenotetext{b}{Full X-ray sample of 44 clusters.} 
\tablenotetext{c}{X-ray subsample of 18 cool-core (CC) clusters.} 
\tablenotetext{d}{X-ray subsample of 7 intermediate (ICC) and 19
non-cool-core (NCC) clusters.}
\tablecomments{The
vector $(a,b,c)$ is the plane normal $P_3$. $\theta$ is the angle
between $P_3$ and the $\log T_X$ axis, and $\phi$ is the azimuthal angle
around the $\log T_X$ axis. The parameters $(r_{s0}, M_{s0}, T_{X0})$
represent the (logarithmic) sample means of $(r_s, M_s, T_X)$.}
\end{deluxetable*}

\begin{deluxetable*}{cccccccc}[h]
\tabletypesize{\footnotesize}
\tablecaption{Cluster X-Ray data\label{tab:data}}
\tablenum{2}
\tablewidth{0pt}
\tablehead{
\colhead{Cluster} & \colhead{$z$} &
\colhead{$r_s$} &
\colhead{$c_{200}$} & \colhead{$M_{\rm 200,XMM}$} & 
\colhead{$M_{\rm 200,Ch}$} & 
\colhead{$T_{\rm XMM}$} & \colhead{$T_{\rm Ch}$}\\
\colhead{} & \colhead{} & \colhead{(kpc)} &
\colhead{} & \colhead{($10^{14}\: M_\odot$) } & 
\colhead{($10^{14}\: M_\odot$) } & \colhead{(keV)} & \colhead{(keV)}
}
\startdata
RXCJ~0003.8+0203  &
  0.092&
$   143
^{+   36}_{-   28}$&
$ 8.06
^{+ 1.52}_{- 1.30}$&
$ 1.90
\pm 0.23$&
$ 2.28
^{+ 0.28}_{- 0.28}$&
$ 4.0
\pm 0.3$&
$ 4.8
^{+ 0.4}_{- 0.4}$
\\
Abell~3911        &
  0.097&
$   261
^{+  108}_{-   59}$&
$ 5.59
^{+ 1.33}_{- 1.39}$&
$ 3.88
\pm 0.50$&
$ 4.77
^{+ 0.62}_{- 0.63}$&
$ 5.1
\pm 0.7$&
$ 6.2
^{+ 1.0}_{- 0.9}$
\\
Abell~3827        &
  0.098&
$   390
^{+   89}_{-   64}$&
$ 4.47
^{+ 0.67}_{- 0.64}$&
$ 6.61
\pm 0.73$&
$ 8.37
^{+ 0.93}_{- 0.93}$&
$ 6.8
\pm 0.3$&
$ 8.6
^{+ 0.4}_{- 0.4}$
\\
RXCJ~0049.4-2931  &
  0.108&
$    71
^{+   30}_{-   19}$&
$12.78
^{+ 3.80}_{- 3.18}$&
$ 0.94
\pm 0.16$&
$ 1.10
^{+ 0.19}_{- 0.19}$&
$ 3.3
\pm 0.8$&
$ 3.8
^{+ 1.0}_{- 1.0}$
\\
Abell~2034        &
  0.113&
$   979
^{+    7}_{-  317}$&
$ 2.46
^{+ 0.81}_{- 0.06}$&
$17.64
\pm 2.17$&
$22.18
^{+ 2.76}_{- 2.78}$&
$ 6.4
\pm 0.9$&
$ 8.0
^{+ 1.3}_{- 1.3}$
\\
RXCJ~1516.5-0056  &
  0.115&
$   563
^{+    0}_{-  114}$&
$ 2.75
^{+ 0.50}_{- 0.06}$&
$ 4.73
\pm 0.42$&
$ 5.66
^{+ 0.51}_{- 0.52}$&
$ 3.9
\pm 0.6$&
$ 4.7
^{+ 0.8}_{- 0.8}$
\\
RXCJ~2149.1-3041  &
  0.118&
$   251
^{+   41}_{-   28}$&
$ 4.79
^{+ 0.43}_{- 0.49}$&
$ 2.21
\pm 0.21$&
$ 2.63
^{+ 0.26}_{- 0.26}$&
$ 3.7
\pm 0.3$&
$ 4.4
^{+ 0.5}_{- 0.4}$
\\
RXCJ~1516.3+0005  &
  0.118&
$   185
^{+   67}_{-   42}$&
$ 7.06
^{+ 1.64}_{- 1.54}$&
$ 2.84
\pm 0.41$&
$ 3.48
^{+ 0.51}_{- 0.51}$&
$ 4.9
\pm 0.2$&
$ 6.1
^{+ 0.4}_{- 0.3}$
\\
RXCJ~1141.4-1216  &
  0.119&
$   496
^{+   60}_{-   36}$&
$ 3.15
^{+ 0.19}_{- 0.24}$&
$ 4.88
\pm 0.37$&
$ 5.76
^{+ 0.45}_{- 0.45}$&
$ 3.5
\pm 0.5$&
$ 4.1
^{+ 0.6}_{- 0.6}$
\\
RXCJ~1044.5-0704  &
  0.132&
$   286
^{+   23}_{-   27}$&
$ 4.56
^{+ 0.34}_{- 0.25}$&
$ 2.86
\pm 0.18$&
$ 3.41
^{+ 0.22}_{- 0.22}$&
$ 3.7
\pm 0.3$&
$ 4.4
^{+ 0.4}_{- 0.4}$
\\
Abell~1068        &
  0.138&
$   564
^{+   66}_{-   49}$&
$ 3.02
^{+ 0.20}_{- 0.22}$&
$ 6.40
\pm 0.48$&
$ 7.73
^{+ 0.61}_{- 0.60}$&
$ 4.3
\pm 0.9$&
$ 5.2
^{+ 1.2}_{- 1.2}$
\\
RXCJ~2218.6-3853  &
  0.138&
$   597
^{+  184}_{-  166}$&
$ 3.16
^{+ 0.85}_{- 0.55}$&
$ 8.76
\pm 1.62$&
$10.98
^{+ 2.03}_{- 2.05}$&
$ 6.2
\pm 0.5$&
$ 7.8
^{+ 0.7}_{- 0.7}$
\\
RXCJ~0605.8-3518  &
  0.141&
$   369
^{+   47}_{-   39}$&
$ 4.10
^{+ 0.34}_{- 0.34}$&
$ 4.51
\pm 0.36$&
$ 5.49
^{+ 0.44}_{- 0.44}$&
$ 4.6
\pm 0.3$&
$ 5.6
^{+ 0.4}_{- 0.4}$
\\
RXCJ~0020.7-2542  &
  0.142&
$   473
^{+  245}_{-  154}$&
$ 4.17
^{+ 1.41}_{- 1.07}$&
$10.03
\pm 2.67$&
$12.39
^{+ 3.31}_{- 3.33}$&
$ 5.5
\pm 1.2$&
$ 6.8
^{+ 1.7}_{- 1.7}$
\\
Abell~1413        &
  0.143&
$   287
^{+   23}_{-   32}$&
$ 5.83
^{+ 0.57}_{- 0.35}$&
$ 6.12
\pm 0.32$&
$ 7.68
^{+ 0.44}_{- 0.43}$&
$ 6.3
\pm 1.1$&
$ 7.9
^{+ 1.5}_{- 1.4}$
\\
RXCJ~2048.1-1750  &
  0.147&
$   742
^{+   80}_{-  370}$&
$ 2.23
^{+ 1.63}_{- 0.21}$&
$ 5.96
\pm 1.12$&
$ 7.34
^{+ 1.38}_{- 1.39}$&
$ 5.2
\pm 0.4$&
$ 6.4
^{+ 0.5}_{- 0.5}$
\\
RXCJ~0547.6-3152  &
  0.148&
$   443
^{+  253}_{-   71}$&
$ 4.10
^{+ 0.59}_{- 1.17}$&
$ 7.89
\pm 1.51$&
$ 9.90
^{+ 1.89}_{- 1.91}$&
$ 6.3
\pm 0.3$&
$ 7.9
^{+ 0.5}_{- 0.5}$
\\
Abell~2204        &
  0.152&
$   816
^{+  137}_{-    0}$&
$ 2.81
^{+ 0.02}_{- 0.28}$&
$15.93
\pm 1.20$&
$20.48
^{+ 1.59}_{- 1.57}$&
$ 8.0
\pm 1.0$&
$10.3
^{+ 1.4}_{- 1.4}$
\\
RXCJ~0958.3-1103  &
  0.153&
$   872
^{+  260}_{-  183}$&
$ 2.39
^{+ 0.42}_{- 0.39}$&
$11.94
\pm 2.02$&
$14.87
^{+ 2.54}_{- 2.55}$&
$ 5.8
\pm 1.0$&
$ 7.3
^{+ 1.5}_{- 1.4}$
\\
RXCJ~2234.5-3744  &
  0.153&
$   506
^{+  261}_{-  220}$&
$ 4.28
^{+ 2.31}_{- 1.16}$&
$13.42
\pm 4.15$&
$17.19
^{+ 5.30}_{- 5.34}$&
$ 7.7
\pm 1.0$&
$ 9.9
^{+ 1.4}_{- 1.4}$
\\
RXCJ~2014.8-2430  &
  0.161&
$   462
^{+   59}_{-   25}$&
$ 3.86
^{+ 0.15}_{- 0.30}$&
$ 7.56
\pm 0.53$&
$ 9.52
^{+ 0.68}_{- 0.68}$&
$ 6.5
\pm 0.6$&
$ 8.2
^{+ 0.8}_{- 0.8}$
\\
RXCJ~0645.4-5413  &
  0.167&
$   380
^{+  135}_{-   89}$&
$ 4.58
^{+ 1.06}_{- 0.96}$&
$ 7.08
\pm 1.12$&
$ 9.14
^{+ 1.45}_{- 1.45}$&
$ 8.4
\pm 0.3$&
$10.8
^{+ 0.5}_{- 0.5}$
\\
Abell~2218        &
  0.176&
$   243
^{+   95}_{-   79}$&
$ 6.26
^{+ 2.46}_{- 1.48}$&
$ 4.76
\pm 0.74$&
$ 5.98
^{+ 0.93}_{- 0.94}$&
$ 6.3
\pm 0.6$&
$ 7.9
^{+ 0.8}_{- 0.8}$
\\
Abell~1689        &
  0.183&
$   211
^{+   22}_{-   19}$&
$ 8.31
^{+ 0.64}_{- 0.63}$&
$ 7.36
\pm 0.44$&
$ 9.52
^{+ 0.59}_{- 0.58}$&
$ 8.5
\pm 0.8$&
$11.0
^{+ 1.2}_{- 1.2}$
\\
Abell~383         &
  0.187&
$   435
^{+   95}_{-    0}$&
$ 3.40
^{+ 0.03}_{- 0.42}$&
$ 4.43
\pm 0.37$&
$ 5.33
^{+ 0.45}_{- 0.46}$&
$ 4.1
\pm 0.3$&
$ 4.9
^{+ 0.4}_{- 0.4}$
\\
Abell~209         &
  0.206&
$   604
^{+  272}_{-  133}$&
$ 3.03
^{+ 0.67}_{- 0.77}$&
$ 8.60
\pm 1.23$&
$10.80
^{+ 1.57}_{- 1.57}$&
$ 6.4
\pm 1.2$&
$ 8.1
^{+ 1.7}_{- 1.6}$
\\
Abell~963         &
  0.206&
$   377
^{+  107}_{-   83}$&
$ 4.35
^{+ 0.94}_{- 0.76}$&
$ 6.17
\pm 0.83$&
$ 7.74
^{+ 1.04}_{- 1.05}$&
$ 6.2
\pm 0.4$&
$ 7.8
^{+ 0.6}_{- 0.6}$
\\
Abell~773         &
  0.217&
$   605
^{+  408}_{-  233}$&
$ 3.27
^{+ 1.49}_{- 1.05}$&
$10.94
\pm 3.12$&
$13.92
^{+ 3.96}_{- 3.98}$&
$ 7.3
\pm 1.0$&
$ 9.3
^{+ 1.4}_{- 1.4}$
\\
Abell~1763        &
  0.223&
$   192
^{+  194}_{-   49}$&
$ 7.50
^{+ 2.30}_{- 3.42}$&
$ 4.25
\pm 0.74$&
$ 5.39
^{+ 0.94}_{- 0.94}$&
$ 6.8
\pm 0.4$&
$ 8.7
^{+ 0.5}_{- 0.5}$
\\
Abell~2390        &
  0.228&
$  1258
^{+    0}_{-   95}$&
$ 2.06
^{+ 0.12}_{- 0.04}$&
$24.71
\pm 1.16$&
$32.55
^{+ 1.89}_{- 1.89}$&
$10.4
\pm 2.8$&
$13.8
^{+ 4.1}_{- 4.0}$
\\
Abell~2667        &
  0.230&
$   993
^{+    0}_{-   48}$&
$ 2.25
^{+ 0.08}_{- 0.02}$&
$15.88
\pm 0.45$&
$20.19
^{+ 0.72}_{- 0.70}$&
$ 7.1
\pm 1.0$&
$ 9.1
^{+ 1.5}_{- 1.4}$
\\
RXCJ~2129.6+0005  &
  0.235&
$   418
^{+   68}_{-   37}$&
$ 3.71
^{+ 0.27}_{- 0.38}$&
$ 5.40
\pm 0.44$&
$ 6.66
^{+ 0.55}_{- 0.55}$&
$ 5.2
\pm 0.5$&
$ 6.5
^{+ 0.7}_{- 0.7}$
\\
Abell~1835        &
  0.253&
$   866
^{+   46}_{-  143}$&
$ 2.64
^{+ 0.34}_{- 0.09}$&
$17.53
\pm 1.41$&
$22.59
^{+ 1.91}_{- 1.88}$&
$ 8.2
\pm 1.5$&
$10.6
^{+ 2.1}_{- 2.1}$
\\
RXCJ~0307.0-2840  &
  0.253&
$   611
^{+  297}_{-  175}$&
$ 3.15
^{+ 0.88}_{- 0.78}$&
$10.44
\pm 2.39$&
$13.03
^{+ 3.01}_{- 3.01}$&
$ 6.1
\pm 1.5$&
$ 7.7
^{+ 2.1}_{- 2.1}$
\\
Abell~68          &
  0.255&
$   834
^{+    0}_{-  257}$&
$ 2.65
^{+ 0.82}_{- 0.06}$&
$15.96
\pm 1.98$&
$20.20
^{+ 2.53}_{- 2.55}$&
$ 6.9
\pm 1.1$&
$ 8.7
^{+ 1.6}_{- 1.5}$
\\
E~1455+2232       &
  0.258&
$   214
^{+   26}_{-   22}$&
$ 6.33
^{+ 0.53}_{- 0.51}$&
$ 3.66
\pm 0.29$&
$ 4.47
^{+ 0.36}_{- 0.36}$&
$ 4.7
\pm 0.5$&
$ 5.7
^{+ 0.7}_{- 0.7}$
\\
RXCJ~2337.6+0016  &
  0.273&
$   332
^{+  342}_{-  154}$&
$ 4.99
^{+ 3.52}_{- 2.18}$&
$ 6.81
\pm 1.91$&
$ 8.61
^{+ 2.41}_{- 2.43}$&
$ 6.7
\pm 1.1$&
$ 8.5
^{+ 1.6}_{- 1.5}$
\\
RXCJ~0303.8-7752  &
  0.274&
$  1115
^{+   14}_{-  497}$&
$ 1.85
^{+ 1.04}_{- 0.09}$&
$13.21
\pm 2.33$&
$16.40
^{+ 3.27}_{- 3.44}$&
$ 7.0
\pm 4.1$&
$ 8.9
^{+ 5.9}_{- 5.5}$
\\
RXCJ~0532.9-3701  &
  0.275&
$   278
^{+  170}_{-   98}$&
$ 5.97
^{+ 2.43}_{- 1.82}$&
$ 6.88
\pm 1.83$&
$ 8.71
^{+ 2.33}_{- 2.34}$&
$ 7.0
\pm 1.6$&
$ 8.9
^{+ 2.3}_{- 2.2}$
\\
RXCJ~0232.4-4420  &
  0.284&
$  1172
^{+    0}_{-  409}$&
$ 1.80
^{+ 0.66}_{- 0.04}$&
$14.28
\pm 1.91$&
$18.37
^{+ 2.48}_{- 2.49}$&
$ 8.1
\pm 1.4$&
$10.4
^{+ 2.0}_{- 2.0}$
\\
ZW~3146           &
  0.291&
$   510
^{+   61}_{-   31}$&
$ 3.37
^{+ 0.15}_{- 0.25}$&
$ 7.79
\pm 0.49$&
$ 9.85
^{+ 0.63}_{- 0.63}$&
$ 6.8
\pm 0.5$&
$ 8.5
^{+ 0.7}_{- 0.7}$
\\
RXCJ~0043.4-2037  &
  0.292&
$   186
^{+  196}_{-   81}$&
$ 7.80
^{+ 5.05}_{- 3.51}$&
$ 4.70
\pm 1.24$&
$ 5.79
^{+ 1.55}_{- 1.55}$&
$ 5.5
\pm 1.6$&
$ 6.9
^{+ 2.3}_{- 2.2}$
\\
RXCJ~0516.7-5430  &
  0.295&
$   785
^{+  405}_{-  472}$&
$ 2.41
^{+ 2.82}_{- 0.75}$&
$10.44
\pm 2.88$&
$13.02
^{+ 3.60}_{- 3.62}$&
$ 5.9
\pm 1.1$&
$ 7.4
^{+ 1.5}_{- 1.4}$
\\
RXCJ~1131.9-1955  &
  0.307&
$   797
^{+  494}_{-  309}$&
$ 2.43
^{+ 1.16}_{- 0.76}$&
$11.31
\pm 2.50$&
$14.06
^{+ 3.23}_{- 3.23}$&
$ 6.3
\pm 2.6$&
$ 8.0
^{+ 3.6}_{- 3.5}$
\\

\enddata
\tablecomments{$M_{\rm 200,XMM}$ is the mass $M_{200}$
originally obtained with {\it XMM-Newton}, and $M_{\rm 200,Ch}$ is the
one corrected for the systematic difference of measured temperature
between {\it Chandra} and {\it XMM-Newton}. $T_{\rm XMM}$ is the X-ray
temperature originally obtained with {\it XMM-Newton}, and $T_{\rm Ch}$
is the corresponding {\it Chandra} temperature
(equation~(\ref{eq:Tch})).}
\end{deluxetable*}

\clearpage

\appendix

\section{The mass ratio between the FP}
\label{sec:app}

First, we assume that the XFP is shifted from the CFP solely in the
direction of $M_s$. From equation~(\ref{eq:MNFW}),
\begin{equation}
\label{eq:Ms}
 M_s = 4\pi\delta_c\rho_c r_s^3(\ln 2 - 1/2)\:.
\end{equation}
Thus, the ratio $f_{Ms}$ is equivalent to that of $\delta_c$ among the
XFP and the CFP or $\delta_{cX}/\delta_{cC}$ if $r_s$ does not depend on
the FPs. If we define $\delta'_c\equiv \delta_c/\Delta$, we have
$f_{Ms}=\delta'_{cX}/\delta'_{cC}$. From equations~(\ref{eq:delc})
and~(\ref{eq:y}), the inverse function of $\delta'_c$ can be defined,
which we call $\tilde{c}_\Delta (\delta'_c)$. From
equations~(\ref{eq:rD}) and~(\ref{eq:c}),
\begin{equation}
\label{eq:fMD}
 f_{M\Delta}= \frac{M_{\Delta X}}{M_{\Delta C}}
=\left(\frac{r_{\Delta X}}{r_{\Delta C}}\right)^3
=\left(\frac{c_{\Delta X}}{c_{\Delta C}}\right)^3
=\left(\frac{c_{\Delta X}}{\tilde{c}_\Delta (\delta'_{cC})}\right)^3
=\left(\frac{c_{\Delta X}}
{\tilde{c}_\Delta (f_{Ms}^{-1}\delta'_{cX})}\right)^3
\end{equation}
Since $\delta'_{cX} = y(c_{\Delta X})$ from equations~(\ref{eq:delc})
and~(\ref{eq:y}), we obtain
\begin{equation}
\label{eq:fMD-cDX}
 f_{M\Delta}
=\left(\frac{c_{\Delta X}}
{\tilde{c}_\Delta (f_{Ms}^{-1}y(c_{\Delta X}))}\right)^3\:,
\end{equation}
which is a function of $c_{\Delta X}$ for a given $f_{Ms}$. It can
also be written as
\begin{equation}
\label{eq:fMD-cDC}
 f_{M\Delta}
=\left(\frac{\tilde{c}_\Delta (f_{Ms}\delta'_{cC})}
{c_{\Delta C}}\right)^3
=\left(\frac{\tilde{c}_\Delta (f_{Ms} y(c_{\Delta C}))}
{c_{\Delta C}}\right)^3\:,
\end{equation}
and it is a function of $c_{\Delta C}$ for a given $f_{Ms}$.

Second, we assume that the XFP is shifted from the CFP solely in the
direction of $r_s$. From equation~(\ref{eq:Ms}), we obtain
$\delta'_{cC}/\delta'_{cX}=\delta_{cC}/\delta_{cX}=(r_{sX}/r_{sC})^3
=f_{rs}^3$ if $M_s$ does not depend on the FPs. From
equations~(\ref{eq:rD}) and~(\ref{eq:c}),
\begin{equation}
\label{eq:fMD2}
 f_{M\Delta}
=\left(\frac{r_{\Delta X}}{r_{\Delta C}}\right)^3
=\left(\frac{c_{\Delta X}r_{sX}}{c_{\Delta C}r_{sC}}\right)^3
=f_{rs}^3\left(\frac{c_{\Delta X}}
{\tilde{c}_\Delta (\delta'_{cC})}\right)^3
=f_{rs}^3\left(\frac{c_{\Delta X}}
{\tilde{c}_\Delta (f_{rs}^3\delta'_{cX})}\right)^3
=f_{rs}^3\left(\frac{c_{\Delta X}}
{\tilde{c}_\Delta (f_{rs}^3 y(c_{\Delta X}))}\right)^3\:,
\end{equation}
which a function of $c_{\Delta X}$ for a given $f_{rs}$. Similarly, we
have
\begin{equation}
\label{eq:fMD-cDC2}
 f_{M\Delta}
=f_{rs}^3\left(\frac{\tilde{c}_\Delta (f_{rs}^{-3}\delta'_{cC})}
{c_{\Delta C}}\right)^3
=f_{rs}^3\left(\frac{\tilde{c}_\Delta (f_{rs}^{-3}y(c_{\Delta C}))}
{c_{\Delta C}}\right)^3\:,
\end{equation}
and it is a function of $c_{\Delta C}$ for a given $f_{rs}$.

\end{document}